\documentclass[12pt]{article}  

\usepackage[a4paper,textwidth=490.0pt,textheight=703.1pt]{geometry}
\usepackage{cite}
\usepackage{graphicx}
\usepackage{epsfig}
\usepackage{amsmath}
\usepackage{amssymb}
\usepackage{ulem}
\usepackage{mdwlist}
\usepackage{color}
\usepackage{float}
\usepackage[rflt]{floatflt}
\usepackage{slashed}

\usepackage{array}

\usepackage[english]{babel}
\usepackage[latin1]{inputenc}
\usepackage[T1]{fontenc}
\usepackage{ae}

\usepackage{url}

\usepackage{amsmath, amsthm, amssymb}

\newtheorem{thm}{Theorem}[section]

\usepackage{array}

\usepackage[english]{babel}
\usepackage[latin1]{inputenc}
\usepackage[T1]{fontenc}
\usepackage{ae}

\newtheorem{definition}[thm]{Definition}

\newcommand{\gsim}{\raisebox{-0.07cm   }
{$\, \stackrel{>}{{\scriptstyle\sim}}\, $}}
\newcommand{\GeV}{\rm GeV}

\newcommand{\ep}{\varepsilon}

\usepackage{rotating}

\usepackage{graphicx}

\newcounter{mmacnt}
\def\restartmma{\setcounter{mmacnt}{0}}
\restartmma \catcode`|=\active
\def|#1|{\mathrm{#1}}
\catcode`|=12
\newenvironment{mma}{
 \par\smallskip
 \catcode`|=\active
 \parskip=0pt\parindent=0pt 
 \small
 \def\In##1\\{%
   \def\linebreak{\hfill\break\null\qquad}%
   \refstepcounter{mmacnt}
   \hangindent=2.5em\hangafter=0
   \leavevmode
   \llap{\tiny\sffamily In[\arabic{mmacnt}]:=\kern.5em}%
   \mathversion{bold}\footnotesize$\displaystyle##1$\normalsize
   \mathversion{normal}\par
 }%
 \def\Print##1\\{%
   \def\linebreak{\hfill\break}%
   \hangindent=2.5em\hangafter=0
   \leavevmode ##1\par}%
 \def\Out##1\\{%
   \def\linebreak{$\hfill\break\null\hfill$}%
   \kern\abovedisplayskip\par
   \hangindent=2.5em\hangafter=0
   \leavevmode
   \llap{\tiny\sffamily Out[\arabic{mmacnt}]=\kern.5em}
   \footnotesize$\displaystyle##1$\normalsize\hfill\null\par
   \kern\belowdisplayskip
 }%
 \def\Warning##1##2\\{%
   \def\linebreak{\hfill\break}%
   \hangindent=2.5em\hangafter=0
   \leavevmode
   {\scriptsize##1 : ##2}\par}%
}{%
 \par\smallskip
}

\usepackage{color}

\newenvironment{fshaded}{%
\MakeFramed {\FrameRestore}
}%
{\endMakeFramed}

\usepackage{tikz}
\usetikzlibrary{matrix}

\allowdisplaybreaks[4]

\begin{document}
\setlength{\baselineskip}{0.515cm}
\sloppy
\thispagestyle{empty}
\begin{flushleft}
DESY 14--098
\hfill 
\\
DO--TH 14/11\\
MITP/15-030\\
April 2015\\
\end{flushleft}

\mbox{}
\vspace*{\fill}
\begin{center}

{\LARGE\bf The 3-Loop Non-Singlet Heavy Flavor}

\vspace*{2mm}
{\LARGE\bf Contributions to the Structure Function}

\vspace*{2mm}
{\LARGE\bf \boldmath $g_1(x,Q^2)$  at Large Momentum Transfer}

\vspace{4cm}
\large
A.~Behring$^a$,
J.~Bl\"umlein$^a$,
A.~De Freitas$^{a,b}$, 
A.~von~Manteuffel$^c$,

\vspace*{1.5mm}
and
C.~Schneider$^b$

\vspace{1.5cm}
\normalsize   
{\it  $^a$~Deutsches Elektronen--Synchrotron, DESY,}\\
{\it  Platanenallee 6, D-15738 Zeuthen, Germany}

\vspace*{3mm}
{\it $^b$~Research Institute for Symbolic Computation (RISC),\\
                          Johannes Kepler University, Altenbergerstra\ss{}e 69,
                          A--4040, Linz, Austria}\\

\vspace*{3mm}   
{\it  $^c$ PRISMA Cluster of Excellence,  Institute of Physics, J. Gutenberg
University,}\\
{\it D-55099 Mainz, Germany.}
\\

\end{center}
\normalsize
\vspace{\fill}
\begin{abstract}
\noindent 
We calculate the massive flavor non-singlet Wilson coefficient for the heavy flavor contributions to 
the polarized structure function $g_1(x,Q^2)$ in the asymptotic region $Q^2 \gg m^2$ to 3-loop order 
in Quantum Chromodynamics at general values of the Mellin variable $N$ and the momentum fraction $x$, 
and derive heavy flavor corrections to the Bjorken sum-rule. Numerical results are presented for the 
charm quark contribution. Results on the structure function $g_2(x,Q^2)$ in the twist-2 approximation
are also given. 
\end{abstract}

\vspace*{\fill}
\noindent
\numberwithin{equation}{section}

\newpage
\section{Introduction}
\label{sec:1}

\vspace*{1mm}
\noindent
Massless and massive contributions to the unpolarized and polarized structure functions in deep-inelastic scattering 
exhibit different scaling violations. For a precise determination of the QCD scale $\Lambda_{\rm QCD}$ or the strong 
coupling constant $\alpha_s(M_Z^2)$ their precise knowledge is therefore of importance \cite{Bethke:2011tr}.
In the case of the polarized structure function $g_1(x,Q^2)$ the complete heavy flavor corrections are only available at 
1-loop order  \cite{Watson:1981ce,Vogelsang:1990ug}\footnote{For an implementation in Mellin space, see \cite{Alekhin:2003ev}.}. 
At higher orders in the coupling constant, the heavy flavor contributions were calculated in the asymptotic region 
$Q^2 \gg m^2$ based on the factorization derived in Ref.~\cite{Buza:1995ie}. Here $Q^2$ denotes the virtuality of the 
exchanged gauge boson and $m$ the heavy quark mass. The $O(\alpha_s^2)$ corrections in the polarized case were 
calculated in Refs.~\cite{Buza:1996xr,Bierenbaum:2007pn}. In the case of the structure function $g_1(x,Q^2)$, the 1-loop 
heavy flavor corrections have been accounted for at next-to-leading order (NLO) QCD analysis \cite{Blumlein:2010rn}. 
The corresponding flavor non-singlet corrections in the unpolarized case were calculated for pure photon exchange to 
$O(\alpha_s^2)$ in \cite{Buza:1995ie,Bierenbaum:2007qe} and in Ref.~\cite{Ablinger:2014vwa} to $O(\alpha_s^3)$. 

In the present paper we calculate the $O(\alpha_s^3)$ massive flavor non-singlet Wilson coefficient for the inclusive structure
function $g_1(x,Q^2)$ in the asymptotic region $Q^2 \gg m^2$, and also present the corresponding $O(\alpha_s^2)$ result, extending
Refs.~\cite{Buza:1996xr,Bierenbaum:2007pn}.

The differential cross section for polarized deep-inelastic scattering \cite{Bardin:1996ch,Lampe:1998eu,Blumlein:2012bf}
is given by
\begin{eqnarray}
\frac{d^2 \sigma_{\rm B}}{dx dy}  = \frac{2 \pi \alpha^2}{Q^4} \lambda_N^p f^p S \left[
  S_1^p(x,y) g_1(x,Q^2)
+ S_2^p(x,y) g_2(x,Q^2) \right],
\end{eqnarray}
with
\begin{eqnarray}
f^L &=& 1,~~~~~~f^T = \cos(\beta - \varphi) \frac{d \varphi}{2 \pi} \sqrt{ \frac{4M^2 x}{Sy} \left[1 - 
y - \frac{M^2 x y}{S} \right]}, 
\nonumber\\
S_1^L &=& 2xy \left[(2-y) - 2 \frac{M^2}{S} xy\right],~~~~~~S_1^T = 2xy^2,
\nonumber\\
S_2^L &=& - 8 x^2 y \frac{M^2}{S},~~~~~~~S_2^T = 4xy~.
\end{eqnarray}
Here $\alpha = e^2/(4\pi)$ denotes the fine structure constant, $M$ is the nucleon mass, $S = (p+l)^2$
is the center of mass energy of the lepton-nucleon system, with $p$ and $l$ the nucleon and lepton 4-momenta, respectively, $q = l-l'$ is
the 4-momentum transfer and $Q^2 = -q^2$.  $x = Q^2/(2p.q)$ and $y = p.q/p.l$ are the Bjorken variables. $\lambda_N^p$ 
denotes the degree of the nucleon polarization. The spin 4--vectors in the longitudinal and transverse cases are given by
\begin{eqnarray}
S_L &=& M (0, 0, 0; 1) \\
S_T &=& M (0,\cos(\beta),\sin(\beta); 0)~,
\end{eqnarray}
and $\varphi$ denotes the angle between the vectors of the spin and the outgoing lepton.
It contributes in a non-trivial way in the case of transverse polarization.

The polarized structure functions are denoted by $g_1(x,Q^2)$ and $g_2(x,Q^2)$. In the leading twist approximation,
the heavy flavor contributions to the structure function $g_1(x,Q^2)$ is given by, cf.~\cite{Buza:1996wv},
\begin{eqnarray}
g_1(x,Q^2) &=& \frac{1}{2} \Biggl\{
\sum_{k=1}^{N_F} e_i^k \Biggl\{ L_{q,g_1}^{\rm NS}\left(x, N_F+1, \frac{Q^2}{m^2},
\frac{m^2}{\mu^2} \right)
\otimes \left[\Delta f_k(x, \mu^2, N_F) + \Delta f_{\bar{k}}(x, \mu^2, N_F) \right]
\nonumber\\ 
& & + \frac{1}{N_F} L_{q,g_1}^{\rm PS}\left(x, N_F+1, \frac{Q^2}{m^2}, \frac{m^2}{\mu^2}\right) \otimes \Delta \Sigma(x,\mu^2, N_F)
\nonumber\\ 
& & + \frac{1}{N_F} L_{g,g_1}^{\rm S}\left(x, N_F+1, \frac{Q^2}{m^2}, \frac{m^2}{\mu^2}\right) \otimes \Delta G(x,\mu^2, N_F) \Biggr\}
\nonumber\\ 
& & + e_Q^2 \Biggl[ H_{q,g_1}^{\rm PS}\left(x, N_F+1, \frac{Q^2}{m^2}, \frac{m^2}{\mu^2}\right) \otimes \Delta \Sigma(x,\mu^2, N_F) 
\nonumber\\ 
& & + H_{g,g_1}^{\rm PS}\left(x, N_F+1, \frac{Q^2}{m^2}, \frac{m^2}{\mu^2}\right) 
\otimes \Delta G(x,\mu^2, N_F) \Biggr]\Biggr\}~,
\label{eq:G1}
\end{eqnarray}
with $\Delta f_{k(\bar{k})}$ the $N_F$ light flavor polarized (anti)quark densities,  $\Delta G$ and $\Delta \Sigma = \sum_{l=1}^{N_F}
[\Delta f_k + \Delta f_{\bar{k}}]$ the polarized gluon and singlet distributions, and $e_i$ and $e_Q$ the electric charges of the light
quarks and the heavy quark $Q$, respectively. $\mu$ denotes the factorization scale and $\otimes$ the Mellin convolution
\begin{eqnarray}
A(x) \otimes B(x) = \int_0^1 dx_1 \int_0^1 dx_2 \delta(x - x_1 x_2) A(x_1) B(x_2)~.
\end{eqnarray}
The actual flavor non-singlet distribution is defined by
\begin{eqnarray}
\Delta^{\rm NS}(x,Q^2) = \sum_{k=1}^{N_F} e_k^2 \left[\Delta f_k(x, \mu^2, N_F) + \Delta f_{\bar{k}}(x, \mu^2, N_F) - 
\frac{1}{N_F} \Delta \Sigma(x,\mu^2, N_F) \right]~.
\end{eqnarray}
However, according to the representation (\ref{eq:G1}), 
we will consider its whole first term, depending on 
$L_{q,1}^{\rm NS}$ as the non-singlet contribution in what follows. The structure function $g_2(x,Q^2)$ can be obtained 
from $g_1(x,Q^2)$ using the Wandzura-Wilson relation \cite{Wandzura:1977qf}.

The paper is organized as follows. In Section~\ref{sec:2} we calculate the heavy flavor contributions to the 
non-singlet Wilson coefficient in the asymptotic region $Q^2 \gg m^2$ to the structure function $g_1(x,Q^2)$ 
to 3-loop order in the strong coupling constant. We present the results both in  Mellin $N$ and $x$--space.
Numerical results are given in Section~\ref{sec:3}. Consequences for the polarized Bjorken sum rule are discussed in
Section~\ref{sec:4}, and Section~\ref{sec:5} contains the conclusions. 
\section{The Wilson Coefficient}
\label{sec:2}

\vspace*{1mm}
\noindent
The heavy flavor non-singlet Wilson coefficient contributing to the structure function $g_1(x,Q^2)$
in the asymptotic region $Q^2 \gg m^2$ receives its first contributions at $O(\alpha_s^2)$. In previous analyses
\cite{Buza:1996xr,Bierenbaum:2007pn} the tagged flavor case at $O(\alpha_s^2)$ has been considered. In what follows 
we will refer to the inclusive case, i.e. the complete contribution to the structure function 
$g_1(x,Q^2)$, and consider the terms due a single heavy quark. 

The non-singlet heavy flavor Wilson coefficient contributing to the structure function $g_1(x,Q^2)$
in the asymptotic region $Q^2 \gg m^2$ is given by \cite{Bierenbaum:2009mv}
\begin{eqnarray}
L_{q,g_1}^{\rm h, NS}(N_F+1) &=& a_s^2\Biggl[A_{qq,Q}^{(2),\rm NS}(N_F+1) + \hat{C}_{q,g_1}^{(2),\rm NS}(N_F)\Biggr]   
\\
&+& a_s^3\Biggl[
  A_{qq,Q}^{(3),\rm NS}(N_F+1)
+ A_{qq,Q}^{(2),\rm NS}(N_F+1) {C}_{q,g_1}^{(1),\rm NS}(N_F+1)
+ \hat{C}_{q,g_1}^{(3),\rm NS}(N_F) \Biggr].
\nonumber
\label{eq:LQ2}
\end{eqnarray}
Here $A_{qq,Q}^{\rm NS}$ is the massive  non-singlet operator matrix element (OME) and the label `$N_F+1$' symbolically denotes that the 
OME is 
calculated at $N_F$ massless and one massive flavor, $a_s = \alpha_s/(4\pi) \equiv g_s^2/(4 \pi)^2$ 
parameterizes the strong coupling constant, and we use the convention
\begin{eqnarray}
\hat{f}(N_F) =  f(N_F+1) - f(N_F)~.
\label{eq:HAT}
\end{eqnarray}
The calculation of the different contributions to the Wilson coefficient is performed in $D = 4 + \ep$ dimensions to regulate
the Feynman integrals. In the present polarized case the treatment of $\gamma_5$ has to be considered. In the flavor non-singlet case
both for the massive OMEs and the massless Wilson coefficients $\gamma_5$ always appears in traces along one massless line and there is a 
Ward-Takahashi identity which implies the use of anti-commuting $\gamma_5$. 

The inclusive massive OME  $A_{qq,Q}^{\rm NS}$ to 3-loop order for even and odd moments $N$ has been calculated
in Ref.~\cite{Ablinger:2014vwa}. The corresponding diagrams have been reduced using integration-by-parts relations
\cite{IBP} applying an extension of the package {\tt Reduze\;2} \cite{REDUZE2}\footnote{The package {\tt Reduze\;2} 
uses the packages {\tt Fermat} \cite{FERMAT} and {\tt Ginac} \cite{Bauer:2000cp}.}. The master integrals have been calculated
using hypergeometric, Mellin-Barnes and differential equation techniques, mapping them to recurrences, which have been solved
by modern summation technologies using extensively the packages {\tt Sigma} \cite{SIG1,SIG2}, {\tt EvaluateMultiSums}, 
{\tt SumProduction} \cite{EMSSP}, {\tt $\rho$sum} \cite{RHOSUM}, and {\tt HarmonicSums} \cite{HARMONICSUMS}.
 
The massless Wilson coefficients $C_{q,g_1}(x,Q^2)$ from 1- to 3-loop order were calculated in 
Refs.~\cite{Furmanski:1981cw,vanNeerven:1991nn,Zijlstra:1993sh,Moch:2008fj}. At 3-loop order those 
of the structure function $g_1$ are obtained by that of $F_3$ \cite{Moch:2008fj}, setting the $d_{abc}$ 
terms in $\hat{C}_{q,g_1}^{(3),\rm NS}(N_F)$ to zero, cf. also \cite{Larin:1991tj,Baikov:2010je}.
The non-singlet OMEs $A_{qq,Q}^{(k), \rm NS}$ at 2- and 3-loop order  were calculated in 
\cite{Buza:1995ie,Bierenbaum:2007qe} and \cite{Ablinger:2014vwa}, respectively.

For comparison, the massless flavor non-singlet Wilson coefficient in Mellin space is given by \cite{Zijlstra:1993sh,Moch:2008fj}
\begin{eqnarray}
L_{q,g_1}^{\rm l,NS}(N_F) &=& 1 + \sum_{k=1}^3 
a_s^k {C}_{q,g_1}^{(k),\rm NS}(N_F)~.
\label{eq:LQ3}
\end{eqnarray}

In Mellin $N$ space the Wilson coefficient can be expressed by nested harmonic sums
$S_{\vec{a}}(N)$ \cite{HSUM} which are defined by
\begin{eqnarray}
S_{b,\vec{a}}(N) = \sum_{k=1}^N \frac{({\rm sign}(b))^k}{k^{|b|}} S_{\vec{a}}(k),~~~S_\emptyset = 1,~b, a_i 
\in
\mathbb{Z}, b, a_i \neq 0, N > 0, N \in \mathbb{N}.
\end{eqnarray}
In the following, we drop the argument $N$ of the harmonic sums and use the short-hand
notation $S_{\vec{a}}(N) \equiv S_{\vec{a}}$. The Wilson coefficients depend on the logarithms
\begin{eqnarray}
L_Q = \ln\left(\frac{Q^2}{\mu^2}\right)~~~~~~~\text{and}~~~~~~L_M = \ln\left(\frac{m^2}{\mu^2}\right),
\end{eqnarray}
where the renormalization scale has been set equal to the factorization scale $\mu = \mu_R = \mu_F$. 

As a short-hand notation we define the leading order splitting function $\Delta \gamma_{qq}^{(0)}$  up to its color factor 
\begin{eqnarray}
\Delta \gamma_{qq}^{(0)} = 4 \left[2 S_1 - \frac{3 N^2+3 N+2}{2 N (N+1)}\right]~.
\end{eqnarray}
The massive Wilson coefficient for the structure function $g_1(x,Q^2)$ in the asymptotic region in 
Mellin space in the on-shell scheme is given by

Here the color factors are given by $C_A = N_c, C_F = (N_c^2-1)/(2 N_c), T_F = 1/2$ in $SU(N_c)$, 
and $N_c = 3$ in the case of Quantum Chromodynamics. $\hat{C}_{q,g_1}^{\mathrm{NS},(3)}(N_F)$ denotes the massless Wilson 
coefficient at 3-loop order, cf. (\ref{eq:HAT}), and the polynomials $P_i$ are given by
\begin{eqnarray}
P_1    &=& -3 N^4-6 N^3-47 N^2-20 N+12 \\
P_2    &=& 7 N^4+14 N^3+3 N^2-4 N-4 \\
P_3    &=& 19 N^4+38 N^3-9 N^2-20 N+4 \\
P_4    &=& 28 N^4+56 N^3+28 N^2+2 N+1 \\
P_5    &=& 33 N^4+54 N^3+9 N^2-52 N-28 \\
P_6    &=& 57 N^4+96 N^3+65 N^2-10 N-24 \\
P_7    &=& 112 N^4+224 N^3+121 N^2+9 N+9 \\
P_8    &=& 141 N^4+246 N^3+241 N^2-8 N-84 \\
P_9    &=& 181 N^4+266 N^3+82 N^2-3 N+18 \\
P_{10} &=& 235 N^4+524 N^3+211 N^2+30 N+72 \\
P_{11} &=& 359 N^4+772 N^3+335 N^2+30 N+72 \\
P_{12} &=& 501 N^4+894 N^3+541 N^2-116 N-204 \\
P_{13} &=& 561 N^4+1122 N^3+767 N^2+302 N+48 \\
P_{14} &=& 1131 N^4+2118 N^3+1307 N^2+32 N-276 \\
P_{15} &=& 1139 N^4+2710 N^3+635 N^2+216 N+828 \\
P_{16} &=& 1199 N^4+2398 N^3+1181 N^2+18 N+90 \\
P_{17} &=& 1220 N^4+2359 N^3+1934 N^2+357 N-138 \\
P_{18} &=& 3 N^5+11 N^4+10 N^3+3 N^2+7 N+8 \\
P_{19} &=& 12 N^5+16 N^4+18 N^3-15 N^2-5 N-8 \\
P_{20} &=& 27 N^5+863 N^4+1573 N^3+1151 N^2+144 N-36 \\
P_{21} &=& 51 N^5+102 N^4+121 N^3+118 N^2+48 N+48 \\
P_{22} &=& 648 N^5-2103 N^4-4278 N^3-3505 N^2-682 N-432 \\
P_{23} &=& 1407 N^5+2418 N^4+1793 N^3+134 N^2-384 N+144 \\
P_{24} &=& -11145 N^6-32355 N^5-37523 N^4-14329 N^3+2392 N^2+120 N-1512 \\
P_{25} &=& -151 N^6-469 N^5-181 N^4+305 N^3+208 N^2+40 N+8 \\
P_{26} &=& 3 N^6+9 N^5+70 N^4+77 N^3+39 N^2-10 N-12 \\
P_{27} &=& 6 N^6+18 N^5-N^4-20 N^3+46 N^2+29 N-6 \\
P_{28} &=& 15 N^6+36 N^5+30 N^4+8 N^3+3 N^2+16 N+20 \\
P_{29} &=& 155 N^6+465 N^5+465 N^4+371 N^3+108 N^2+108 N+54 \\
P_{30} &=& 216 N^6+567 N^5+687 N^4+381 N^3+37 N^2-44 N+12 \\
P_{31} &=& 309 N^6+807 N^5+693 N^4-271 N^3-638 N^2+68 N+216 \\
P_{32} &=& 525 N^6+1575 N^5+1535 N^4+973 N^3+536 N^2+48 N-72 \\
P_{33} &=& 609 N^6+1485 N^5+1393 N^4+83 N^3-422 N^2+156 N+216 \\
P_{34} &=& 795 N^6+2043 N^5+2075 N^4+517 N^3-298 N^2+156 N+216 \\
P_{35} &=& 868 N^6+2469 N^5+2487 N^4+940 N^3+27 N^2+63 N+72 \\
P_{36} &=& 1770 N^6+4671 N^5+4765 N^4+1205 N^3-227 N^2+1044 N+756 \\
P_{37} &=& 7531 N^6+23673 N^5+23055 N^4+7375 N^3+1614 N^2+936 N-324 \\
P_{38} &=& -4785 N^7-14355 N^6-4399 N^5+10327 N^4+3548 N^3+3000 N^2
        \nonumber \\ && 
        +1080 N-1728 \\
P_{39} &=& 25 N^7+138 N^6+311 N^5+464 N^4+672 N^3+670 N^2+264 N+48 \\
P_{40} &=& -45 N^8-162 N^7-858 N^6-1960 N^5-1885 N^4-1094 N^3-804 N^2
        \nonumber \\ && 
        -40 N+192 \\
P_{41} &=& 39 N^8+138 N^7+847 N^6+1371 N^5+1283 N^4+485 N^3+101 N^2
\nonumber\\ &&
+132 N+72 \\
P_{42} &=& 3549 N^8+14196 N^7+23870 N^6+25380 N^5+15165 N^4+1712 N^3-2016 N^2
        \nonumber \\ && 
        +144 N+432 \\
P_{43} &=& 5487 N^8+21948 N^7+36370 N^6+28836 N^5+11943 N^4+4312 N^3+2016 N^2
        \nonumber \\ && 
        -144 N-432 \\
P_{44} &=& 10807 N^8+43228 N^7+62898 N^6+39178 N^5+7027 N^4+702 N^3+3240 N^2
        \nonumber \\ && 
        +3456 N+1620 \\
P_{45} &=& 42591 N^8+166764 N^7+245088 N^6+128254 N^5-26735 N^4-40762 N^3
        \nonumber \\ && 
        -3928 N^2-1272 N-2160 \\
P_{46} &=& -18351 N^{10}-89784 N^9-208773 N^8-267222 N^7-192265 N^6-46700 N^5
        \nonumber \\ && 
        +14565 N^4+7730 N^3+1240 N^2+1464 N+144 \\
P_{47} &=& 165 N^{10}+825 N^9+106856 N^8+321746 N^7+396657 N^6+247433 N^5
        \nonumber \\ && 
        +126914 N^4+51804 N^3+6336 N^2+4752 N+5184 \\
P_{48} &=& 828 N^{11}+7632 N^{10}+29217 N^9+59592 N^8+66844 N^7+35738 N^6+7405 N^5
        \nonumber \\ && 
        +16688 N^4+27880 N^3+11552 N^2-3312 N-2304 \\
P_{49} &=& 8274 N^{11}+78519 N^{10}+313841 N^9+686295 N^8+881001 N^7+638778 N^6
        \nonumber \\ && 
        +204948 N^5+7992 N^4+32296 N^3+26544 N^2-10656 N-8640~.
\end{eqnarray}

We would like to note that we disagree with the $O(a_s^2 \ln(Q^2/\mu^2))$ terms given in \cite{Zijlstra:1993sh}, but agree with
the representation in \cite{Moch:2008fj,Moch:2014sna}.

One obtains the analytic continuation of the harmonic sums to complex values of $N$ by performing their asymptotic expansion
analytically, cf.~\cite{Blumlein:2009ta,Blumlein:2009fz}.\footnote{These expansions can now be obtained automatically using the 
package {\tt HarmonicSums} \cite{HARMONICSUMS}.} Furthermore, the nested harmonic sums obey the shift relations 
\begin{eqnarray}
S_{b,\vec{a}}(N) = S_{b,\vec{a}}(N-1) + \frac{{\rm sign}(b)^N}{N^{|b|}} S_{\vec{a}}(N)~, 
\end{eqnarray}
through which any regular point in the complex plane can be reached using the analytic asymptotic representation as input.
The poles of the nested harmonic sums $S_{\vec{a}}(N)$ are located at the non-positive integers. In data analyses,
one may thus encode the QCD evolution \cite{Blumlein:1997em} together with the Wilson coefficient for complex values of $N$
analytically and finally perform one numerical contour integral around the singularities of the problem.\footnote{For precise
numerical implementations of the analytic continuation of harmonic sums see \cite{ANCONT}.}

In $x$-space the Wilson coefficient is represented in terms of harmonic polylogarithms \cite{Remiddi:1999ew} over the alphabet
$\{f_0, f_1, f_{-1}\}$, which were again reduced applying the shuffle relations \cite{Blumlein:2003gb}. They are defined by
\begin{eqnarray}
H_{b,\vec{a}}(x) &=& \int_0^x dy f_b(y) H_{\vec{a}}(y),~~~H_{\scriptsize \underbrace{0,...,0}_{k}}(x) = \frac{1}{k!} 
\ln^k(x),~~~H_\emptyset 
= 1,\\
f_0(x) &=& \frac{1}{x},~~~~~~f_1(x) = \frac{1}{1-x},~~~~~~f_{-1}(x) = \frac{1}{1+x}~.
\end{eqnarray}
The Wilson coefficient is represented by three contributions, the $(...)_+$-function term, the $\delta(1-x)$-term,
and the regular term. Here the $+$-distribution is defined by    
\begin{eqnarray}
\int_0^1 dy \left[F(y)\right]_+ g(y) = \int_0^1 dy F(y) \left[g(y) - g(1)\right]~.
\end{eqnarray}
One obtains 


Again, we used the short hand notation $H_{\vec{a}}(x) \equiv H_{\vec{a}}$ also here.
The transformation of the Wilson coefficient to the $\overline{\rm MS}$ scheme for the heavy quark mass affects the massive 
OME at 3-loops and was given in Ref.~\cite{Ablinger:2014vwa}; the terms are the same in the 
unpolarized and polarized case.

The non-singlet contributions to the structure function $g_2(x,Q^2)$ can be obtained via the Wandzura-Wilczek relation
\cite{Wandzura:1977qf}
\begin{eqnarray}
\label{eq:WW}
g_2(x,Q^2) = -g_1(x,Q^2) + \int_x^1 \frac{dy}{y} g_1(y,Q^2), 
\end{eqnarray}
where both structure functions refer to the twist-2 contributions. This relation is implied by a relation of the OMEs in the 
light-cone expansion, cf.~\cite{Blumlein:1996vs}. The relation has also been proven 
in the covariant parton model in Refs.~\cite{Jackson:1989ph,Roberts:1996ub,Blumlein:1996tp}. For gluonic initial 
states, it was derived in \cite{Blumlein:2003wk}. Eq.~(\ref{eq:WW}) also holds including target mass corrections 
\cite{Piccione:1997zh,Blumlein:1998nv} and finite light quark contributions \cite{Blumlein:1998nv}. Furthermore, it holds in 
non-forward \cite{Blumlein:2000cx} and diffractive  scattering, including target mass corrections 
\cite{Blumlein:2002fw,Blumlein:2008di}.

\section{Numerical Results}
\label{sec:3}

\vspace*{1mm}
\noindent
In what follows, we will choose the factorization and renormalization scale $\mu^2 = Q^2$. We first study the behaviour of the
massive and massless Wilson coefficients in the small and large $x$ region and then give numerical illustrations in the whole $x$-region.
 
At small $x$, the pure massive Wilson coefficient behaves like
\begin{eqnarray}
L_{q,g_1}^{\rm h, NS}(N_F+1) - \hat{C}_{q,g_1}^{\mathrm{NS},(3)}(N_F)
&\propto& a_s^2 4 C_F T_F \ln^2(x) + a_s^3 \Biggl[\frac{16}{27} C_A C_F T_F  - \frac{5}{9} C_F^2 T_F \Biggr] \ln^4(x), 
\end{eqnarray}
while in the region $x \rightarrow 1$ one obtains
\begin{eqnarray}
L_{q,g_1}^{\rm h, NS}(N_F+1) -\hat{C}_{q,g_1}^{\mathrm{NS},(3)}(N_F)
&\propto& 
a_s^2 C_F T_F \frac{8}{3} \left(\frac{\ln^2(1-x)}{1-x}\right)_+
+ a_s^3 C_F^2 T_F \Biggl[16 \ln^2\left(\frac{Q^2}{m^2}\right) 
\nonumber\\
&&
+ \frac{160}{3} \ln^2\left(\frac{Q^2}{m^2}\right) +  
\frac{448}{9}  \Biggr] \left(\frac{\ln^2(1-x)}{1-x}\right)_+~.
\nonumber\\ &&
\end{eqnarray}
There is a term $\propto \ln^3(1-x)/(1-x)$ at $O(\ln(Q^2/\mu^2))$, being of relevance for different choices of the 
factorization scale.

The above results can be compared with the case of the massless Wilson coefficient 
\begin{eqnarray}
\hat{C}_{q,g_1}^{\mathrm{NS},(2)}(N_F) &\propto& a_s^2 \frac{10}{3} C_F T_F \ln^2(x)
\\
\hat{C}_{q,g_1}^{\mathrm{NS},(3)}(N_F)
&\propto& a_s^3 \left[\frac{92}{27} C_F C_A T_F -   
\frac{31}{9} C_F^2 T_F \right] \ln^4(x)
\\                               
\hat{C}_{q,g_1}^{\mathrm{NS},(2)}(N_F) &\propto& a_s^2 \frac{8}{3} C_F T_F \left(\frac{\ln^2(1-x)}{1-x}\right)_+
\\
\hat{C}_{q,g_1}^{\mathrm{NS},(3)}(N_F) &\propto& a_s^3 \frac{80}{9} C_F^2 T_F 
\left(\frac{\ln^4(1-x)}{1-x}\right)_+~.
\end{eqnarray}
The small $x$ behaviour can be compared with leading order predictions for the non-singlet evolution kernel 
in Refs.~\cite{SX,Kiyo:1996si}. Indeed both the massive and massless contributions follow the 
principle pattern $\sim c_k a_s^{k+1} \ln^{2k}(x)$. However, as is well known \cite{SX}, less 
singular terms widely cancel the numerical effect of these leading terms. For the large $x$ terms 
the massless terms exhibit a stronger soft singularity than the massive ones.

In the following numerical illustrations we use the polarized parton distributions of Ref.~\cite{Blumlein:2010rn}, which are 
of next-to-leading order (NLO), since no next-to-next-to-leading order (NNLO) data analysis based on the anomalous dimensions 
calculated in Ref.~\cite{Moch:2014sna} has been performed yet. The values of $\alpha_s$ correspond to those of the unpolarized 
NNLO analysis \cite{Alekhin:2013nda}. The heavy and light flavor Wilson coefficients being discussed in the following
are given in Eqs.~(\ref{eq:LQ2}) and (\ref{eq:LQ3}).

In Figure~1, the 2- and 3-loop heavy flavor corrections to the non-singlet term of the structure function $xg_1(x,Q^2)$ 
are calculated in the case of charm, assuming $m_c = 1.59~\GeV$ \cite{Alekhin:2012vu}, 
using the formula for the Wilson coefficient  Eq.~(\ref{WILx}), and setting $\mu^2 = Q^2$. With growing $Q^2$, the distribution 
diminishes at larger values of $x$ and grows towards medium values. The $O(\alpha_s^3)$ corrections lead to stronger 
effects if compared to those at $O(\alpha_s^2)$. We have applied the asymptotic Wilson coefficients for all
the $Q^2$ values given here, which only holds for values $Q^2/m^2 \gsim 10$. For the heavy quark distributions we formally show
also the result at $Q^2 = 4~\GeV^2$, outside this region, indicated by dotted $(O(a_s^2))$ and dash-dotted lines $(O(a_s^3))$.

Figure~2 shows the effect of the Wilson coefficients 
comparing the contributions from $O(\alpha_s^0)$ to $O(\alpha_s^3)$ at $Q^2 = 4~\GeV^2$ as an example, where a depletion is 
obtained with growing order.
The 3-loop light flavor contributions to $xg_1(x,Q^2)$ ($N_F = 3$) are illustrated in Figure~3. 
Here the evolution is strengthened  by growing $Q^2$ in the large $x$ region and depleted for lower values of $Q^2$, 
considering only the effects due to the Wilson coefficient. 

In Figures~4 and 5 we illustrate the ratio of the flavor non-singlet charm corrections to those by the light quarks given in 
Eq.~(\ref{eq:LQ3}) up to $O(\alpha_s^2)$ and $O(\alpha_s^3)$, respectively. At $O(\alpha_s^2)$ the effect is of $O(1\%)$ and 
below, for the lower scales $Q^2$, but higher values are obtained for very large scales as $Q^2 \simeq 1000~\GeV^2$ in the region
$x \sim 0.003$. A qualitatively
similar picture is obtained including the $O(\alpha_s^3)$ corrections. The effect on the ratio 
$g_1^{\rm heavy}/g_1^{\rm light}|_{\rm NS}$ is about doubled. To resolve relative effects of $O(2 \%)$ requires 
higher luminosities than available in present day experiments. They may become available in the planned experiments
at a future EIC \cite{EIC}.
 
Figure~6 shows the 2- and 3-loop charm flavor non-singlet contributions to the structure function $xg_2(x,Q^2)$ according to the 
Wandzura-Wilczek relation (\ref{eq:WW}) implying the oscillatory behaviour. In size these effects are comparable to those of the 
structure function $xg_1(x,Q^2)$ shown in Figure~1. With growing $Q^2$ the effects become somewhat smaller. In Figure 7 we show
the corresponding massless contributions to the structure function $g_2(x,Q^2)$ at $Q^2 = 4~\GeV^2$ for the different orders
in $a_s$, which slightly diminish adding higher order contributions. Taking into account the $O(a_s^3)$ corrections, the light flavor 
corrections to $g_2(x,Q^2)$ (\ref{eq:G1},\ref{eq:WW}) grow somewhat in size with larger values of $Q^2$, see Figure~8. Similar to 
the case of the 
structure
function $xg_1$ the $O(a_s^3)$ charm flavor non-singlet corrections to the structure function $xg_2(x,Q^2)$ amount to $O(1\%)$.    
\section{The Bjorken Sum Rule}
\label{sec:4}

\vspace*{1mm}
\noindent
The polarized  Bjorken sum rule \cite{Bjorken:1969mm} refers to the first moment of the flavor non-singlet combination
\begin{eqnarray}
\label{eq:PBJ}
\int_0^1 {dx} \left[g_1^{ep}(x,Q^2) - g_1^{en}(x,Q^2) \right] = \frac{1}{6}\left|\frac{g_A}{g_V}\right|
C_{\rm BJ}(\hat{a}_s),
\end{eqnarray}
with $g_{A,V}$ the neutron decay constants, $g_A/g_V \approx -1.2767 \pm 0.0016$ \cite{Mund:2012fq} and $\hat{a}_s = 
\alpha_s/\pi$. The 1- \cite{Kodaira:1978sh}, 2- 
\cite{Gorishnii:1985xm}, 3- \cite{Larin:1991tj} and 4-loop QCD corrections \cite{Baikov:2010je} in the massless 
case are given by
\begin{eqnarray}
\label{eq:BJSR}
C_{\rm BJ}(\hat{a}_s),
&=& 
1 -  \hat{a}_s 
+ \hat{a}_s^2 (-4.58333 + 0.33333 N_F)
+ \hat{a}_s^3 (-41.4399 + 7.60729 N_F - 0.17747 N_F^2)
\nonumber\\ &&
+ \hat{a}_s^4 (-479.448 + 123.472 N_F - 7.69747 N_F^2 + 0.10374 N_F^3)~,
\end{eqnarray} 
choosing the renormalization scale $\mu^2 = Q^2$, cf.~\cite{Zijlstra:1993sh} for $SU(3)_c$. Here $N_F$ denotes the number of 
active light flavors. The expression for general color factors was given in Ref.~\cite{Baikov:2010je}.

For the asymptotic massive corrections (\ref{eq:LQ2}) only the first moments of the massless Wilson coefficients 
$\hat{C}_{q,g_1}^{(2,3),\rm NS}(N_F)$ contribute, since the first moments of the massive non-singlet OMEs vanish due to fermion number 
conservation, a property holding even at higher order. Therefore, any new heavy quark changes Eq.~(\ref{eq:BJSR}) by a shift in 
$N_F
\rightarrow N_F +1$ only, for the asymptotic corrections. Different results are obtained in the tagged flavor case 
\cite{Buza:1995ie,Bierenbaum:2007pn} at $O(\alpha_s^2)$, where no inclusive structure functions are considered. Corresponding 
power corrections were derived in \cite{Blumlein:1998sh,vanNeerven:1999ec}.

\section{Conclusions} 
\label{sec:5} 

\vspace{1mm}
\noindent
We calculated the heavy flavor non-singlet Wilson coefficients of the polarized inclusive structure function $g_1(x,Q^2)$
to $O(\alpha_s^3)$ in the asymptotic region $Q^2 \gg m^2$. The first contributions of this kind are of $O(\alpha_s^2)$.
In the case of twist-2 operators the corresponding contributions to the structure function $g_2(x,Q^2)$ can be obtained
using the Wandzura-Wilczek relation (\ref{eq:WW}) \cite{Wandzura:1977qf}, 
cf.~\cite{Jackson:1989ph,Roberts:1996ub,Blumlein:1996vs,Blumlein:1996tp,Blumlein:1998nv}. 
The asymptotic Wilson coefficient is obtained by using the factorization formula \cite{Buza:1995ie}, Eq. (\ref{eq:LQ2}),
based on the massive OME \cite{Ablinger:2014vwa} and the massless Wilson coefficient \cite{Moch:2008fj} to 3-loop order.
The heavy flavor Wilson coefficient can be thoroughly
represented by nested harmonic sums in Mellin-$N$ space and by harmonic polylogarithms in $x$-space.
We presented numerical results corresponding to the charge weighted polarized parton contributions $\propto \Delta f(x,Q^2) + 
\Delta \bar{f}(x,Q^2)$, cf. (\ref{eq:G1}), referring to the polarized parton distribution functions at NLO \cite{Blumlein:2010rn}
for an illustration. Comparing with  the corresponding massless cases the heavy flavor corrections in case of charm are of 
$O(1-2\%)$, requiring high luminosity experiments to be resolved, which are planned for the future electron-ion collider EIC
\cite{EIC}. We also considered the contribution of the asymptotic Wilson coefficient to the polarized Bjorken sum-rule.  
Due to fermion number conservation for the massive flavor non-singlet OME in all orders in $\alpha_s$, only the first moment of
the massless Wilson coefficient contributes and the effect of each heavy flavor results in a shift of $N_F$ by one unit
in the expression for the massless polarized Bjorken sum-rule.
The results of the present calculation could be easily applied to derive the asymptotic heavy 
flavor corrections to the neutral current structure function $xG_3$, \cite{Arbuzov:1995id}. 
However, the corresponding massless Wilson coefficient to 3-loop order has not been calculated yet.

\vspace{5mm}
\noindent
{\bf Acknowledgment.}~
We would like to thank J.~Ablinger, K.~Chetyrkin, A.~Hasselhuhn, M.~Round, A.~Vogt, 
and F.~Wi\ss{}brock for discussions. This work was supported in part by the European 
Commission through contract PITN-GA-2012-316704 ({HIGGSTOOLS}) and Austrian Science 
Fund (FWF) grant SFB F50 (F5009-N15).

\newpage
\begin{figure}[H]
\centering
\includegraphics[width=0.8\textwidth]{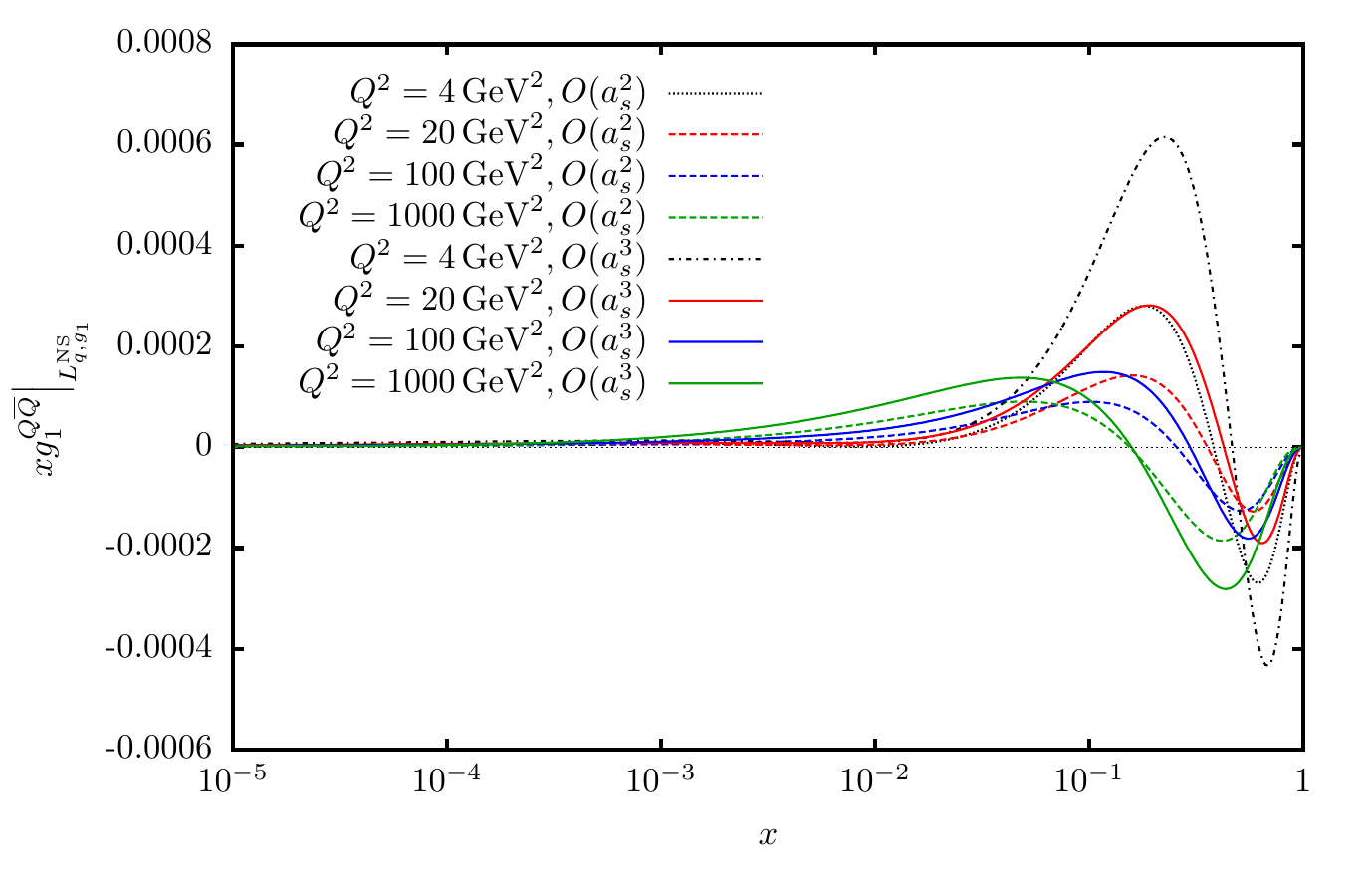}  
\caption{\sf \small The 2- and 3-loop non-singlet charm contributions to the structure function $xg_1(x,Q^2)$ by the
asymptotic heavy flavor Wilson coefficients in the on-shell scheme for $m_c = 1.59~\GeV$. Here we used 
the value of $\alpha_s(M_Z^2) = 0.1132$ and the NLO parton distribution
\cite{Blumlein:2010rn} as reference. Figures~2--8 below are calculated using the same setting.}
\label{fig:1} 
\end{figure}
\begin{figure}[H]
\centering
\includegraphics[width=0.8\textwidth]{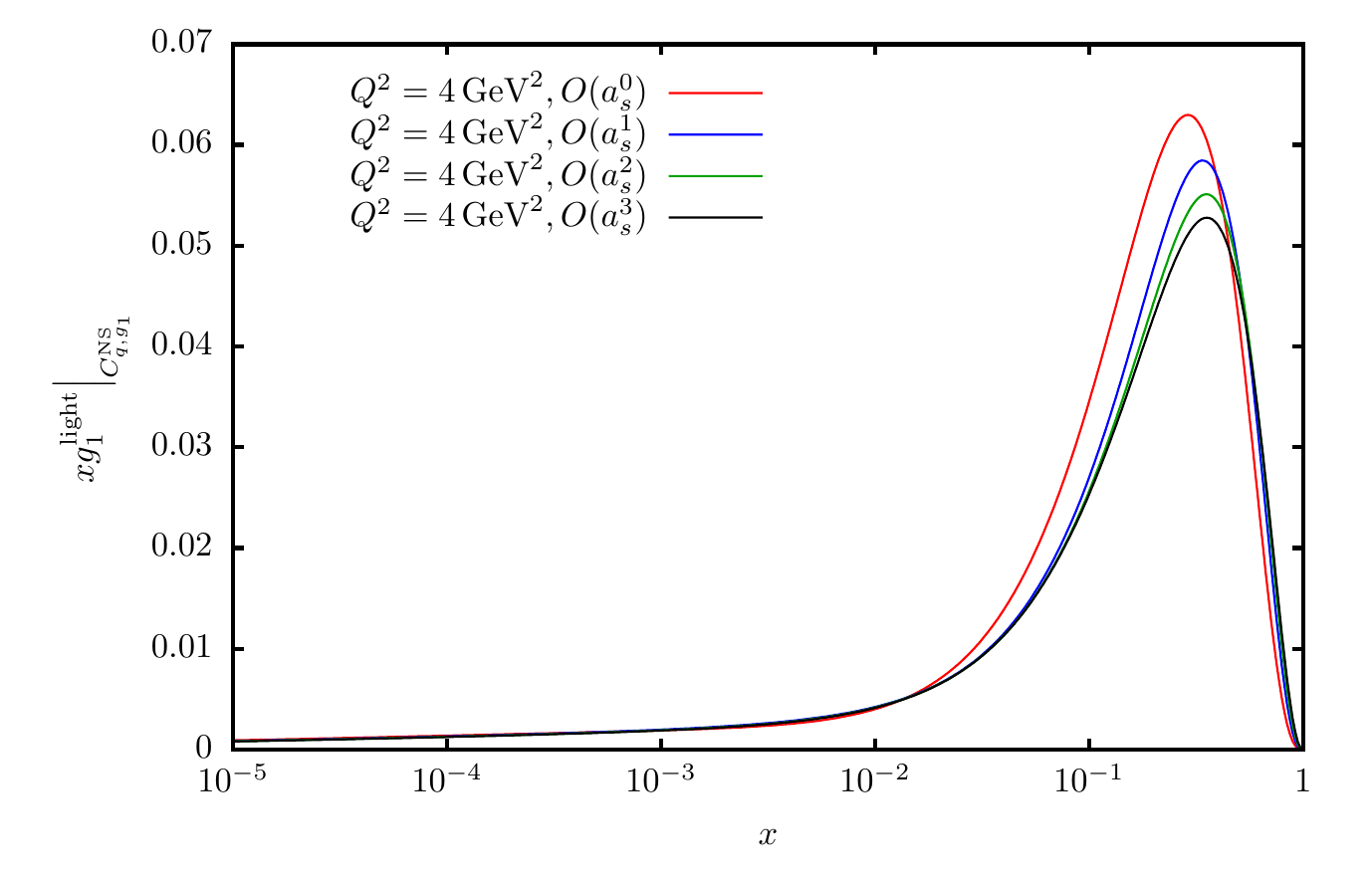}  
\caption{\sf \small The light flavor contributions $(N_F = 3)$ to the non-singlet charm contributions to the structure function 
$xg_1(x,Q^2)$ at $Q^2 = 4~\GeV^2$ illustrating the contributions for the different orders in $a_s$.}
\label{fig:2} 
\end{figure}
\newpage
\begin{figure}[H]
\centering
\includegraphics[width=0.8\textwidth]{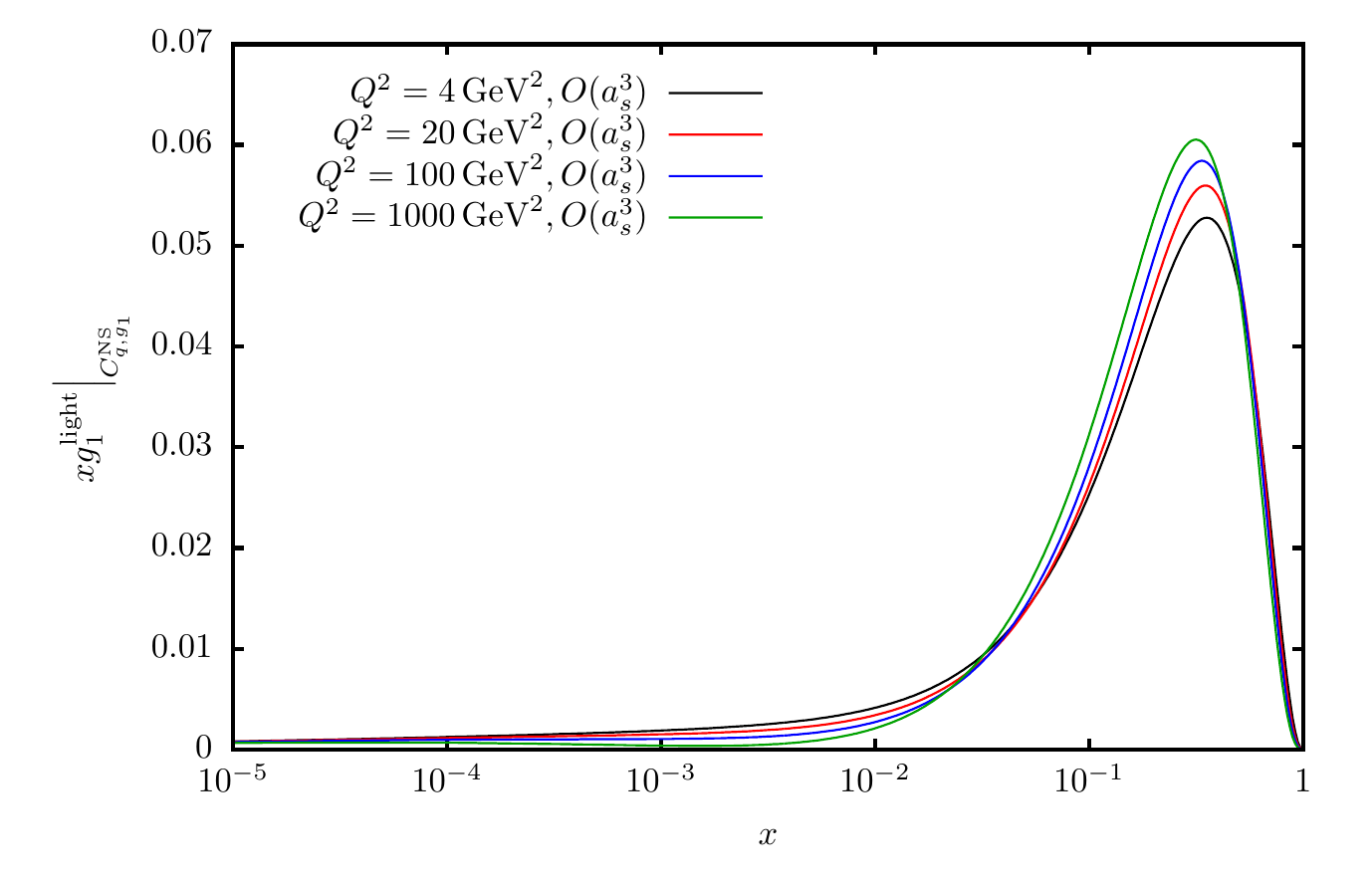}
\caption{\sf \small The light flavor contributions $(N_F = 3)$ to the non-singlet charm contributions to the structure function 
$xg_1(x,Q^2)$ at $O(a_s^3)$ for different values of $Q^2$.}
\label{fig:3} 
\end{figure}
\begin{figure}[H]
\centering
\includegraphics[width=0.8\textwidth]{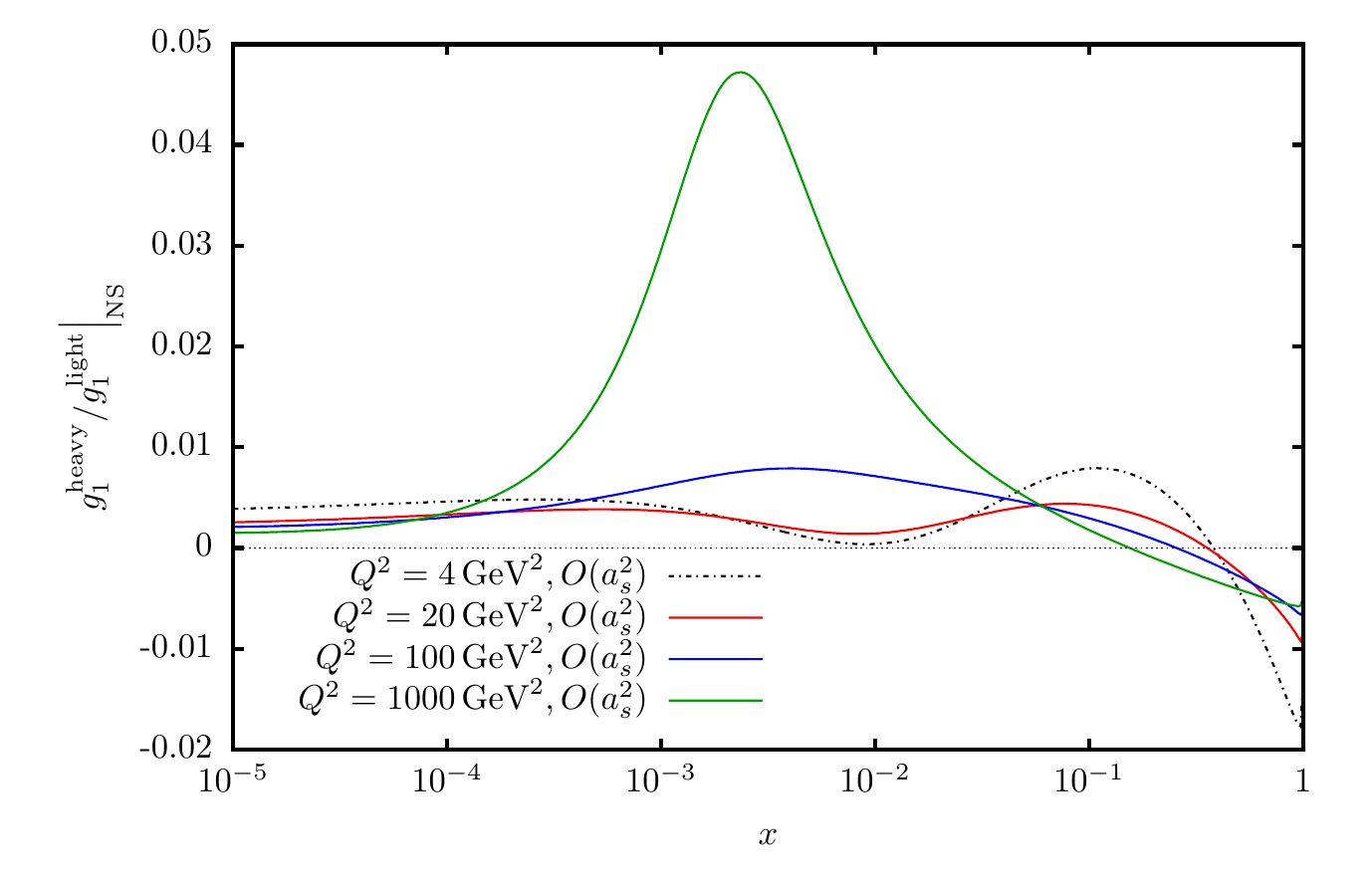}  
\caption{\sf \small The ratio $g_1^{\rm charm}/g_1^{\rm light}|_{\rm NS}$ in the  non-singlet case
at $O(a_s^2)$ for different values of $Q^2$.}
\label{fig:4} 
\end{figure}
\newpage
\begin{figure}[H]
\centering
\includegraphics[width=0.8\textwidth]{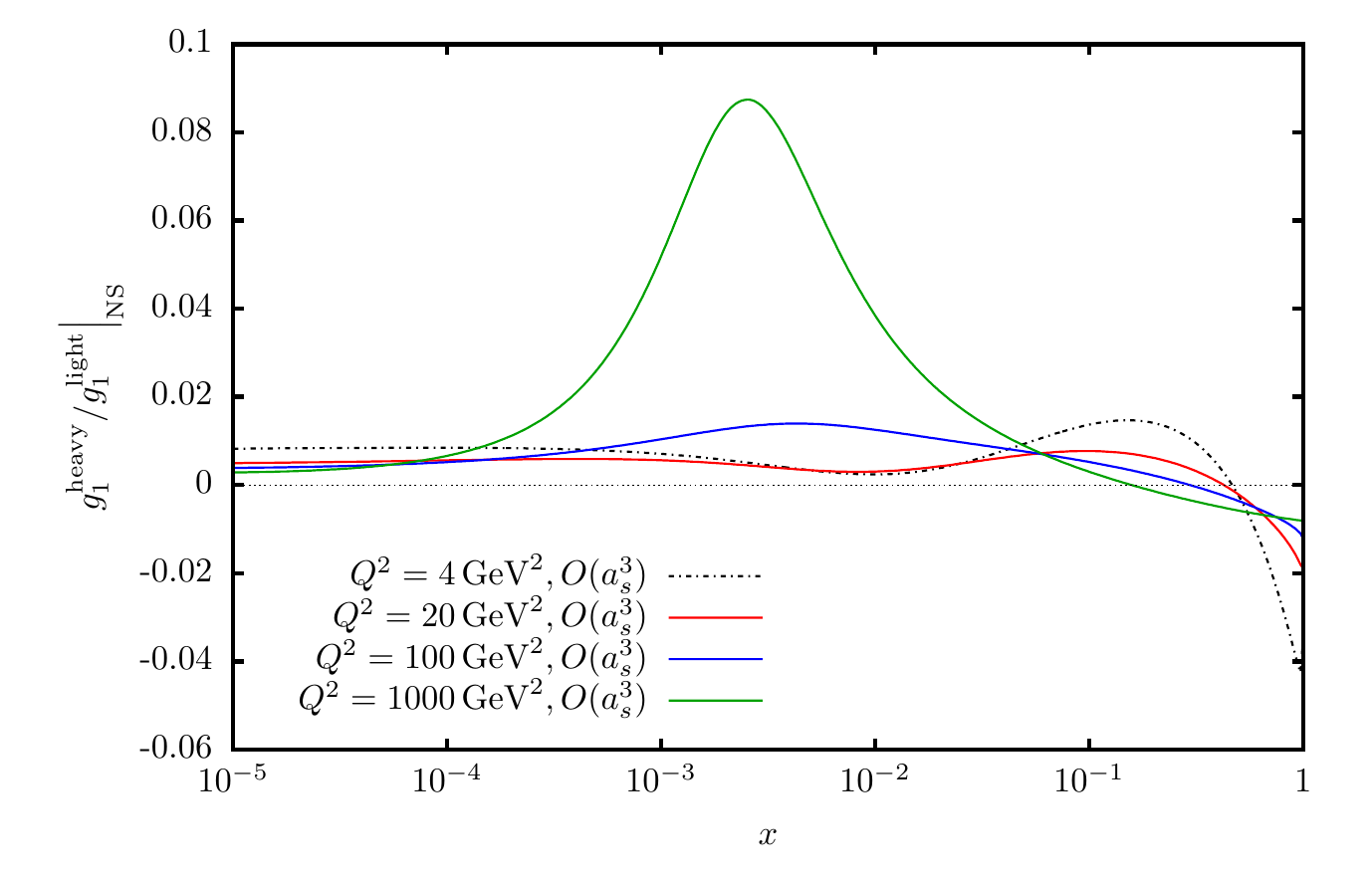}  
\caption{\sf \small The ratio $g_1^{\rm charm}/g_1^{\rm light}|_{\rm NS}$ in the  non-singlet case
at $O(a_s^3)$ for different values of $Q^2$.}
\label{fig:5} 
\end{figure}
\begin{figure}[H]
\centering
\includegraphics[width=0.8\textwidth]{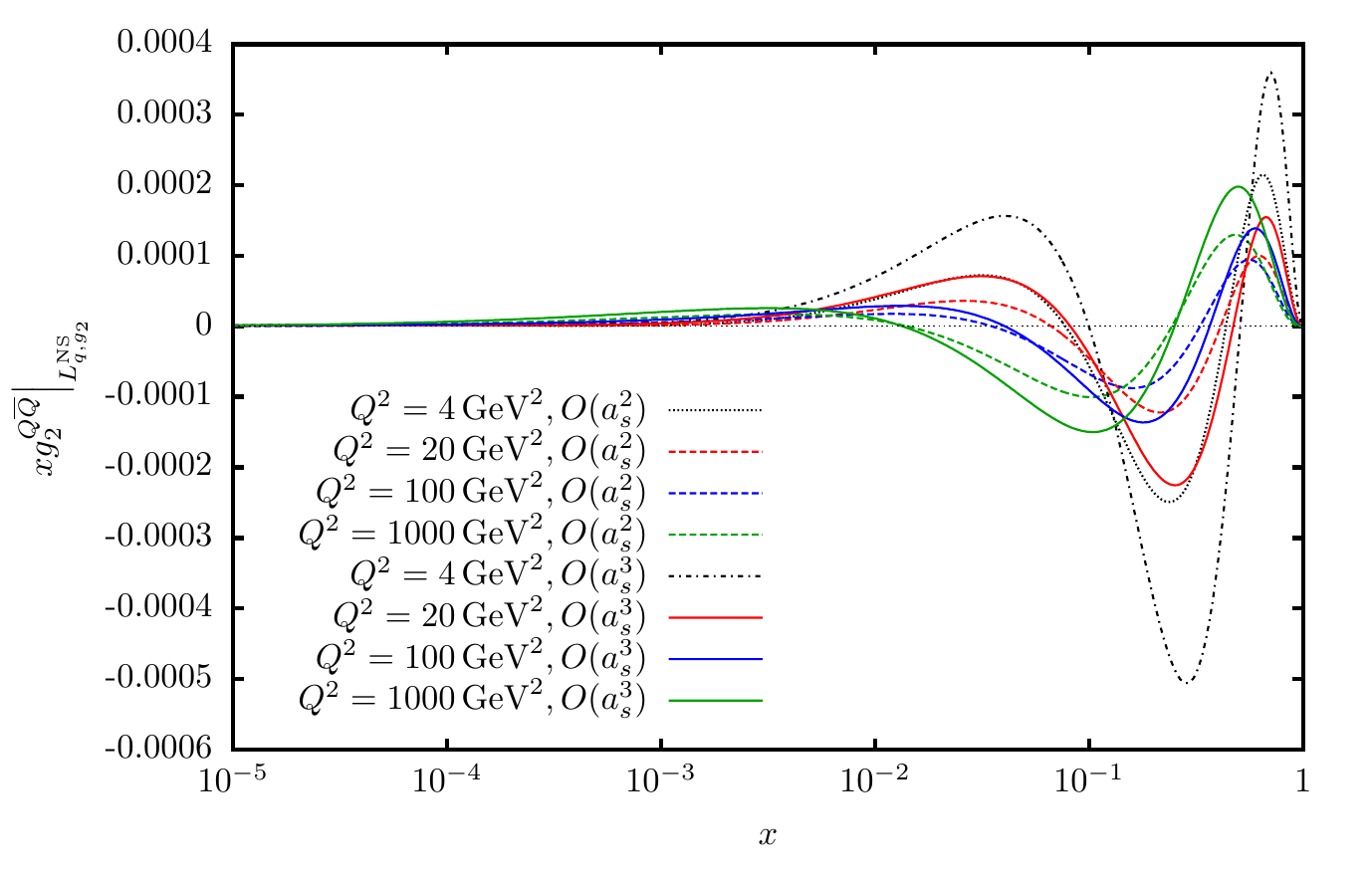}  
\caption{\sf \small The 2- and 3-loop non-singlet charm contributions to the twist 2 contributions of the 
structure function $xg_2(x,Q^2)$ by the asymptotic heavy flavor Wilson coefficients in the on-shell scheme.}
\label{fig:6} 
\end{figure}
\newpage
\begin{figure}[H]
\centering
\includegraphics[width=0.8\textwidth]{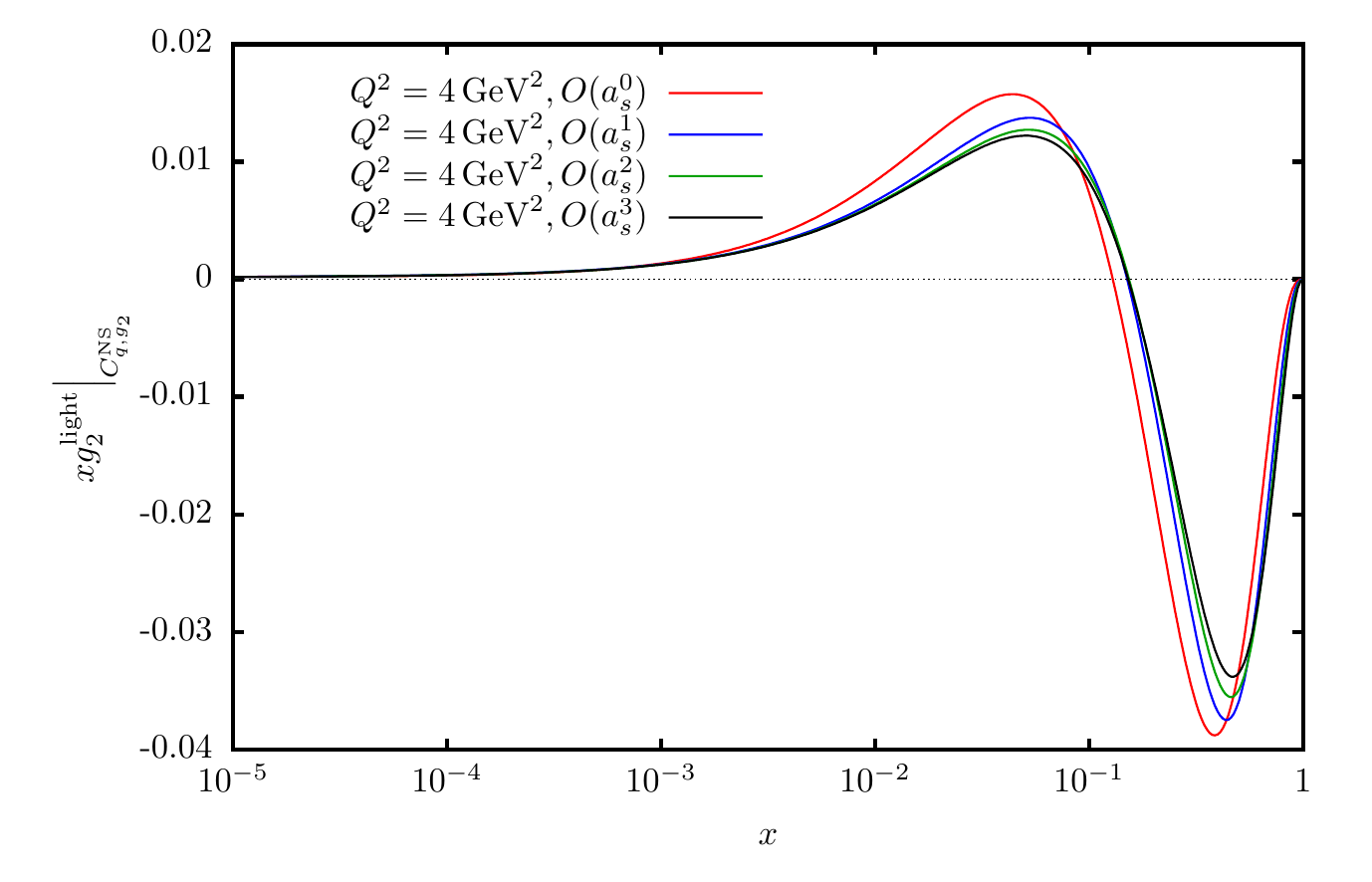}  
\caption{\sf \small The light flavor contributions $(N_F = 3)$ to the non-singlet charm contributions to the structure function 
$xg_2(x,Q^2)$ at $Q^2 = 4~\GeV^2$ illustrating the contributions for the different orders in $a_s$.}
\label{fig:7} 
\end{figure}
\begin{figure}[H]
\centering
\includegraphics[width=0.8\textwidth]{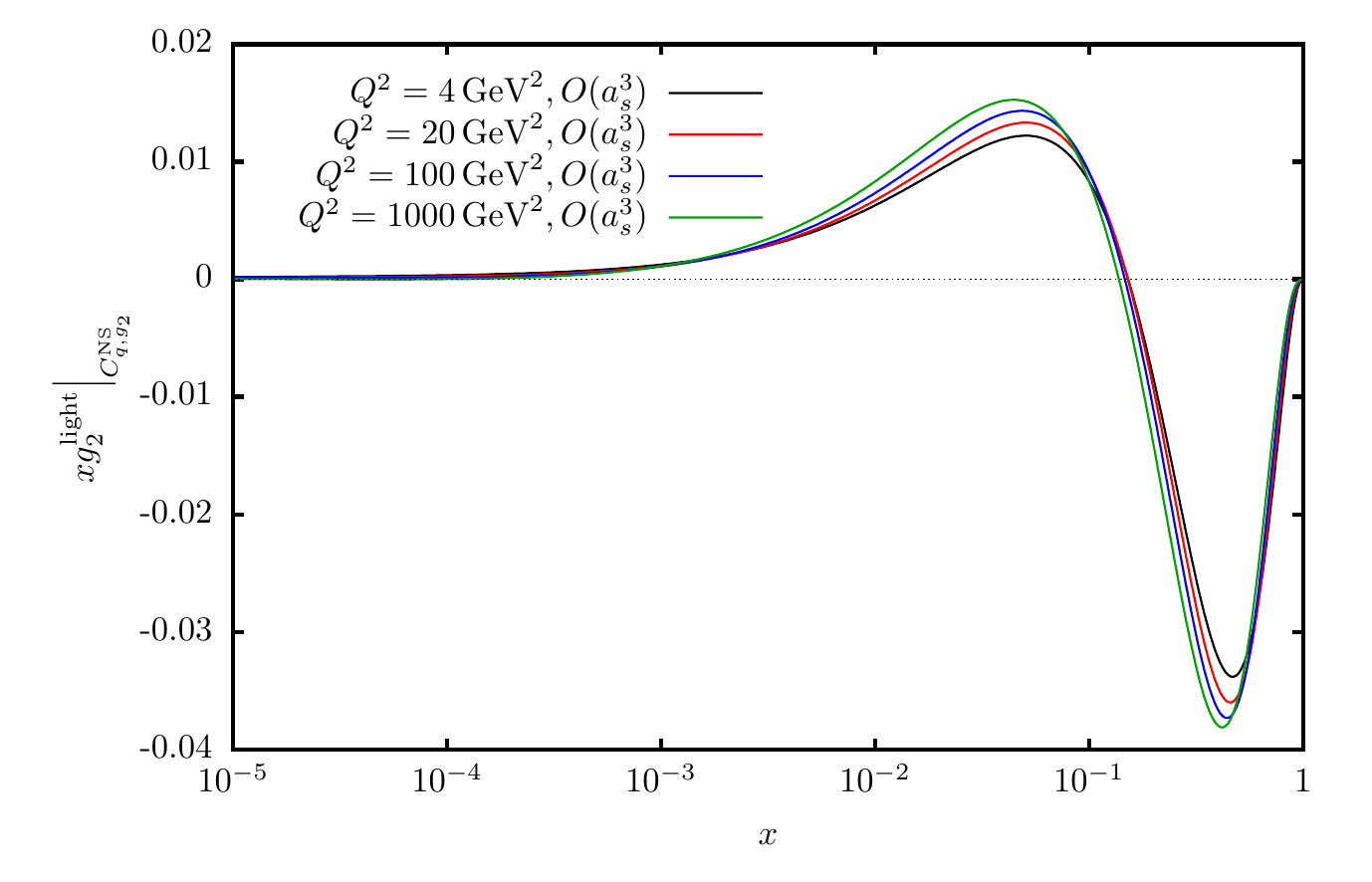}
\caption{\sf \small The light flavor contributions $(N_F = 3)$ to the non-singlet charm contributions to the structure function 
$xg_2(x,Q^2)$ at $O(a_s^3)$ for different values of $Q^2$.}
\label{fig:8} 
\end{figure}

\newpage

\end{document}